\documentclass[manuscript,screen]{acmart}

\AtBeginDocument{%
  \providecommand\BibTeX{{%
    \normalfont B\kern-0.5em{\scshape i\kern-0.25em b}\kern-0.8em\TeX}}}

\begin{document}

\title{A User Interface for Sense-making of the Reasoning Process while Interacting with Robots
}

\author{Chao Wang}
\email{chao.wang@honda-ri.de}
\affiliation{%
  \institution{Honda Research Institute Europe}
  \streetaddress{Carl-Legien-Straße 30}
  \city{Offenbach am Main}
  \state{Hessen}
  \country{Germany}
  \postcode{63073}
}

\author{Joerg Deigmoeller}
\affiliation{%
  \institution{Honda Research Institute Europe}
  \streetaddress{Carl-Legien-Straße 30}
  \city{Offenbach am Main}
  \state{Hessen}
  \country{Germany}
  \postcode{63073}
}

\author{Pengcheng An}
\affiliation{%
  \institution{School of Design, Southern University of Science and Technology}
  \city{Shenzhen}
  \country{China}
}


\author{Julian Eggert}
\affiliation{%
  \institution{Honda Research Institute Europe}
  \streetaddress{Carl-Legien-Straße 30}
  \city{Offenbach am Main}
  \state{Hessen}
  \country{Germany}
  \postcode{63073}
}

\renewcommand{\shortauthors}{Trovato and Tobin, et al.}

\begin{abstract}
This paper describes an interface that enables experts to communicate with a robot via natural language, and visualize the robot's reasoning on commonsense combined with a simulated environment. The interface visually links the robot's internal processes and knowledge with the simulated instances in the form of a 3D isometric visualization as well as the robot's first-person view. After 3 weeks of usage of the system by robotic experts in their daily development, some feedback was collected, which provided insights for designing such systems in the future.


\end{abstract}

\ccsdesc[500]{Human-centered computing~Visualization toolkits}

\keywords{graph representation, human-robot interaction, explainable AI, data visualization}


\received{20 February 2007}
\received[revised]{12 March 2009}
\received[accepted]{5 June 2009}

\maketitle

\section{Introduction}
In the recent years, robots acting in simulated home environments by combining commonsense knowledge and observations, draws more and more attention in the robotics research community. Nevertheless, most of the works usually miss an understanding of the situation and contextual setting, because it requires a well defined representation of an agent's knowledge. One promising form of such a representation are knowledge graphs, which allow and incremental growing of the knowledge during operation by accessing external knowledge sources or through interaction with the user.
Another important advantage of graph representations is their transparency. As graph representations can be human and machine readable at the same time, they provide a meaningful and insightful explanation of how systems reason \cite{Tiddi2022}. Therefore, the visualization of graphs have been investigated extensively for helping researchers and developers to make sense of such data format \cite{Wang2018,Xian2019,Ma2019}. The challenge is still to communicate efficiently and avoid huge amount of information the user has to parse. This requires a good understanding of the current context to focus on most relevant processing steps and data only. There is also a huge potential to use similar visualization methods for helping developers to improve robotic reasoning systems based on graph representation. 
However, this direction is less explored by the previous research, which focuses on human-robot interaction (HRI) and \textit{how} to interact, control, or supervise a robotic system in real-time, but less on data-centric sense-making of \textit{why} the robot makes a decision. However, prior explorations in the domain of eXplainable AI (XAI) and Visualization \cite{Szafr2021} suggest that it is promising to develop interfaces that could support users' sense-making of robots' decisions, especially in real-world environments.

\par

To address this under-explored opportunity, we designed and implemented an interface which allows expert-users to ask questions and send commands to the robot in natural language. The interface shows the decision-making process of the robot, based on the knowledge representation in real-time. Furthermore, users can also modify the graph on the fly if a mistake is found. After several weeks of using the system in the experts' daily development of the robotic reasoning system, insights were summarized based on the feedback of the users. 

In the reminder of this paper, we will first give an overview of related work in Section \ref{sec:related_work}. Then we briefly introduce the back-end (Section \ref{sec:knowledge_engine}), followed by the front-end (Section \ref{sec:interface}) and results of a small user study \ref{sec:feedback}. Finally, we conclude the paper and give an outlook in Section \ref{sec:conlusion}.

\section{related works}\label{sec:related_work}
In the domain of Human-Robot interaction, unfortunately the area of interactive and real-time visualizations for depicting robot decisions on real-world tasks is unfortunately not explored yet. However, there is a large number of interfaces proposed in the HRI domain regarding situational awareness (SA) and control \cite{Szafr2021}. For example, many teleoperating-robot interfaces provide both the map information and video stream from the camera of the robots \cite{Szafr2021}. Some of them even fused the 2D camera image into the 3D map \cite{Nielsen2006,Yanco2007}. In-depth levels of visualization in such interfaces are to include the sensing, perception, prediction, planning and execution information for showing the internal state and goals of the intelligent agents \cite{Govindaraj2013,Szafir2017,Whitlock2020}. 
However, the aim of such interfaces are usually to provide a supervision feature, instead of visualizing the internal reasoning or intervene if necessary in the decision process. \par

In contrast, in the data visualization community, many studies and interfaces have been proposed for sense-making. Especially, the knowledge graph is seen as a tool for enhancing the transparency and explainability of an intelligent system \cite{Tiddi2022, Kawamura2019, LeuceF2020}, which has been applied for many domains, such as visual debugging \cite{Strobelt2019} or recommendation system \cite{Ma2019}. There are many visualization designs proposed for the knowledge graph. For example, Force Simulation \cite{Heer2005}, Force-Directed Edge Bundling \cite{Holten2009} and Arc diagram \cite{Wattenberg2002} etc. However, up to our knowledge, such kind of interfaces have not been investigated yet in the robotic domain for two main reasons: Firstly, visualizations are used for data inspection, instead of interacting with the system by modifying processes or content on the fly. Especially, robotic developers need such a feature, as they require to change the behaviour of the agent immediately. Secondly, it is important to link the abstract knowledge to the concrete instances in the authentic environment where the robot is operating. Therefore, there is a high demand to adapt the knowledge graph for usage by robotic experts.

\section{knowledge representation and system structure}\label{sec:knowledge_engine}
\begin{figure}[h]
\centering
\includegraphics[width=0.9\textwidth]{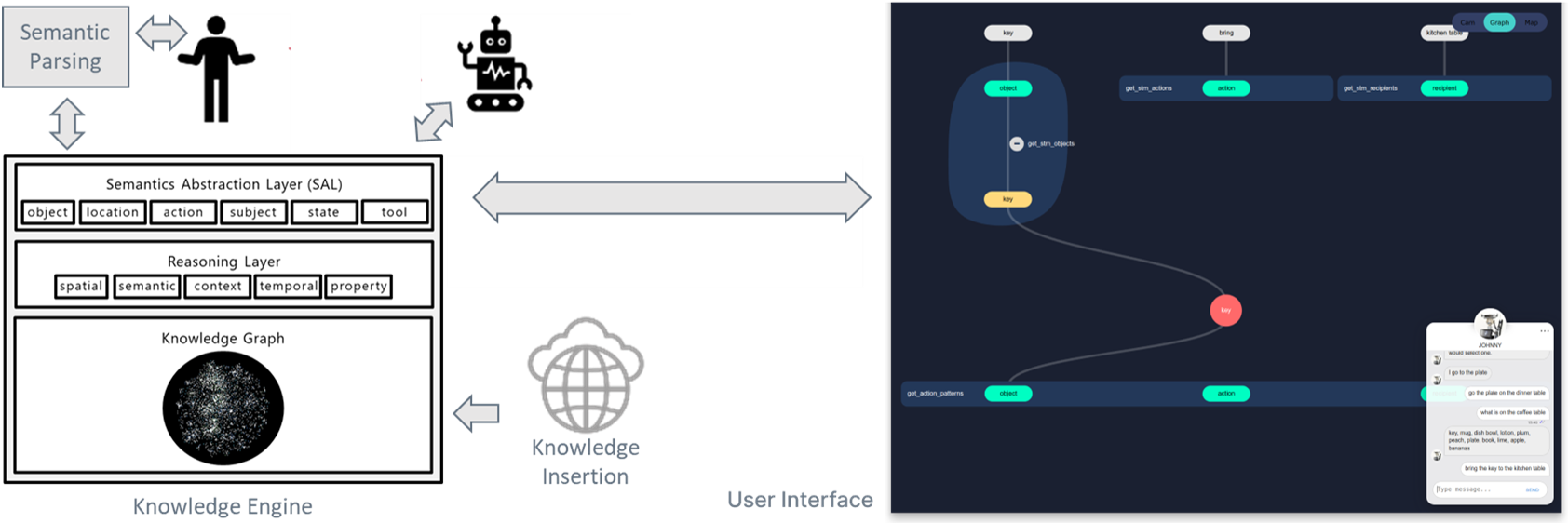}
\caption{Overall system sketch with the Knowledge Engine as core component. The simulated environment and the semantic parsing allows for accessing the Knowledge Engine by natural language. The User Interface facilitates tracing of the reasoning steps and content of the Knowledge Engine.}
\centering
\label{fig:system}
\end{figure}

In this section, we briefly describe the overall system including its components. The core of the system is a Knowledge Engine (KE) that acts as central component (see figure \ref{fig:system}). It consists of a knowledge graph according to MemNet [omitted due to blind review] 
plus reasoning methods operating on the graph. Attached to the KE is a simulator that allows an interaction with a 3D environment, using the knowledge and reasoning capabilities of the KE. The simulator we use is VirtualHome \cite{Puig2018}, where we added a new room environment that is true to scale with our lab. VirtualHome provides an easy interface to control one or multiple agents by the actions walk/run to, walk forward, turn left/right, grab/put, open/close, switch on/off, touch and look at. We use a single robotic agent only, with combined atomic actions for bring (walk to + grab + put) and go (walk to). Further actions are straightforward, nevertheless, have not been implemented at the state of this experiment. The Semantic Parsing allows to translate natural language input (written text in a chat-box) to the previously mentioned actions or giving answers to questions about the environment of type locating ("where is [object]"), counting ("how many [object] are on [location]") or enumerating ("what is on [location]"). The challenge for the KE is to resolve the instances in the environment from indirect references or synonyms in the language input. Additionally, we extracted commonsense knowledge from ConceptNet about objects, locations or tools that can be used for a certain action, as described in [omitted due to blind review]
. This commonsense information is inserted into our KE (see Knowledge Insertion in Figure \ref{fig:system}) as so called \emph{action patterns}. The XAI compnent allows for an intuitive visualization of the internal processes and is the focus in this paper. For the interaction between KE and XAI, it is important to have an understanding of the high level KE access, which is the Semantics Abstraction Layer (SAL), as described in [omitted due to blind review]. 

The SAL is based on getting or setting objects, locations, actions, states (object properties or spatial relations), tools or combinations of those as action patterns. All arguments can be either forwarded as a textual utterance (lemma) or a unique concept ID in the knowledge graph. Further distinction is made between abstract concepts and short term (STM) instances that are attached with properties. We focus on the most important API calls for this paper, which are the STM retrieving functions \emph{get\_stm\_objects, get\_stm\_locations, get\_stm\_actions, get\_count}. The function share the arguments \emph{action, object, location} and \emph{state}. These arguments can be of type lemmas or concept IDs. For each concept, the internal reasoning will also explore its child concepts until they finally hit a matching instance of the environment. That means, we could alternatively identify a \emph{banana} by calling \emph{get\_stm\_objects (object="fruit", state="yellow")} or the location of a \emph{banana} by \emph{get\_stm\_locations (object="fruit", state="yellow")}.

As return value, the functions return always unique concept IDs, which again can be forwarded to any other function, so that we can create a tree of nested calls. Such trees provide an abstract access to the KE reasoning and is the basis for the user interface. 

\begin{figure}[h]
\includegraphics[width=0.8\textwidth]{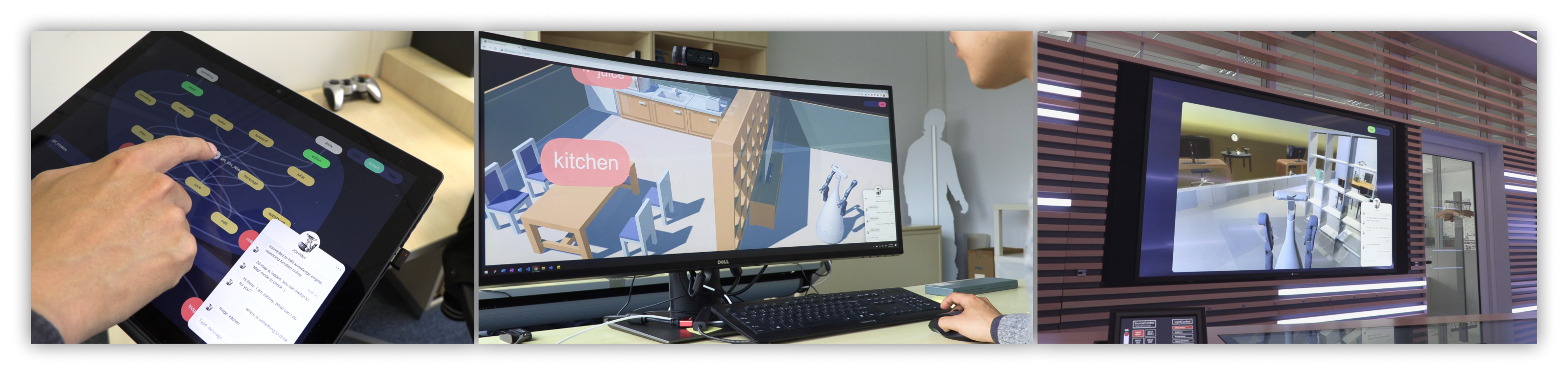}
\caption{User accessing the interface from tablet (left) and PC (middle) or for inspection only on a video panel (right). The interface provides three modes: camera mode (left), map mode (middle) and camera mode (right). On each device, the user can switch between the different modes, as required.}
\centering
\label{fig:scen}
\end{figure}

\section{THE USER Interface}\label{sec:interface}

Before stepping into details of the front-end, we summarize the requirements we like to meet. These requirements are based on feedback of expert users concerning the initial workflow for investigating and developing our system.

\subsection{Expert user requirements}\label{sec:requirements}
In the beginning of the project, the system inspection was quite distributed and uncomfortable. 
The expert user had to open different applications to start and interact with the system: including opening the VirtualHome application to initiate the robotic simulator; open multiple Linux consoles\footnote{https://en.wikipedia.org/wiki/Linux\_console} to start the Knowledge Engine and yet another visualization to crawl the huge knowledge graph. Therefore, we designed and implemented a front-end that combines the distributed system access points into a single web-page, as well as a transparent and efficient visualization of internal decision processes. In a first step, we collected requirements from experts' point of view for a holistic interface:

\begin{enumerate}
    \item sending natural language command via chat-box
    \item trace internal reasoning process and data structure
    \item communicate on basic concepts of the KE, not on raw data level
    \item interface should be accessible from different hosts within a network
    \item modify existing knowledge in the knowledge graph
    \item insert new knowledge via chat-box
\end{enumerate}

In this paper we concentrate on items 1-4, touch a bit item 5 and leave a more detailed study of 5 and 6 for future work.

\subsection{Interface Components}\label{sec:interface}
The graphical user interface (GUI) has been designed and implemented as a web-based front-end. Driven by the back-end, expert users can visit the interface anywhere in the research lab with any device (Figure \ref{fig:scen}). With this we meet already the requirement 4 of our list in Section \ref{sec:requirements}. The chat-box  interface (see Figure \ref{fig:cam-mode}.a) allows the user to type in natural language commands and receive answers from the KE, which is in line with requirement 1. By clicking on the toggle button (see Figure \ref{fig:cam-mode}.b), the user can switch between camera-mode (Figure \ref{fig:cam-mode} left), graph-mode (Figure \ref{fig:graph-mode}) and map-mode (Figure \ref{fig:map-mode}), which provides a holistic approach to inspect the system through different access points. In the following sections, we introduce each more in detail.

\subsubsection{Camera mode}

\begin{figure}[h]
\includegraphics[width=1\textwidth]{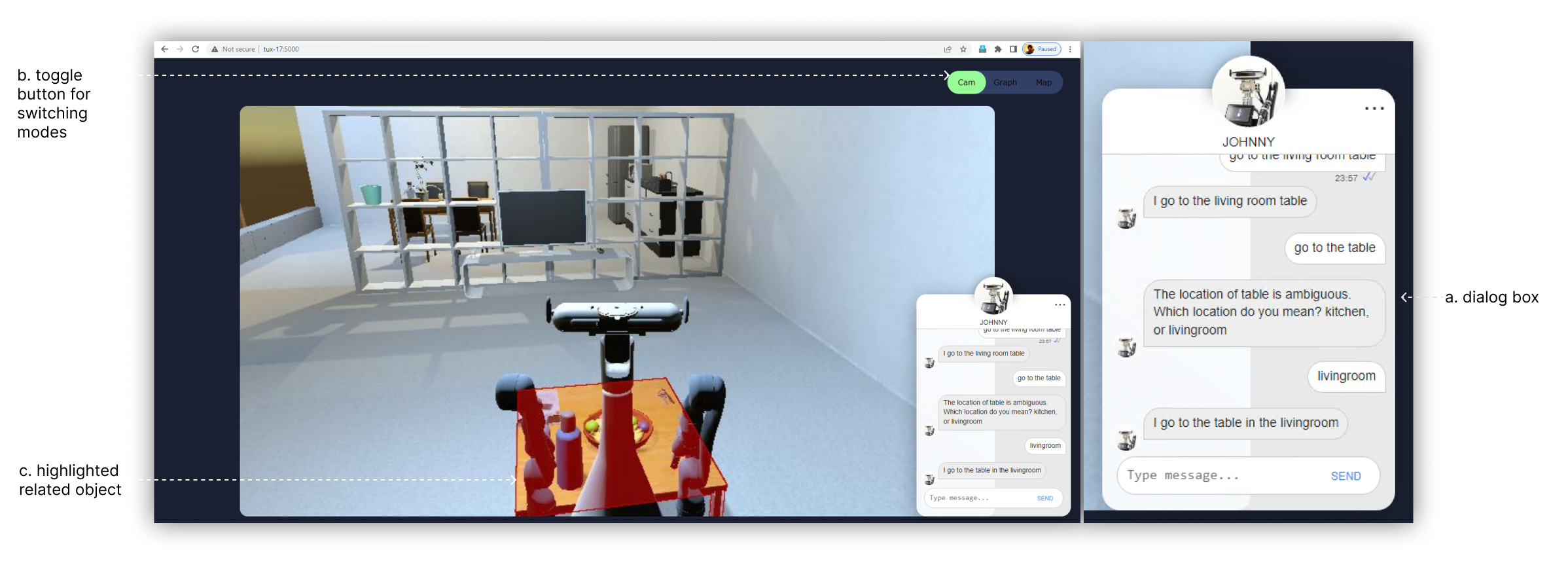}
\caption{The camera mode, showing the simulation in first-person view as video stream (left). Objects are highlighted in the simulator if they are part of the conversation in the chat-box (either by user or answer of the agent) or if the agent considers them for executing an action.}
\centering
\label{fig:cam-mode}
\end{figure}
In the camera mode, the video streaming of the simulator first-person view is displayed (see Figure \ref{fig:cam-mode}, which allows the user to supervise the robot's behavior in real-time. This mode highlights objects in the scene that are part of a conversation in the chat-box. As the example in Figure \ref{fig:cam-mode} shows, after the robot received the command "go to the table", the robot asks the user to specify whether it is the table in the kitchen or the living room, and then it will go to the table. And the table is highlighted.


\subsubsection{Graph Mode}
\begin{figure}[h]
\includegraphics[width=1\textwidth]{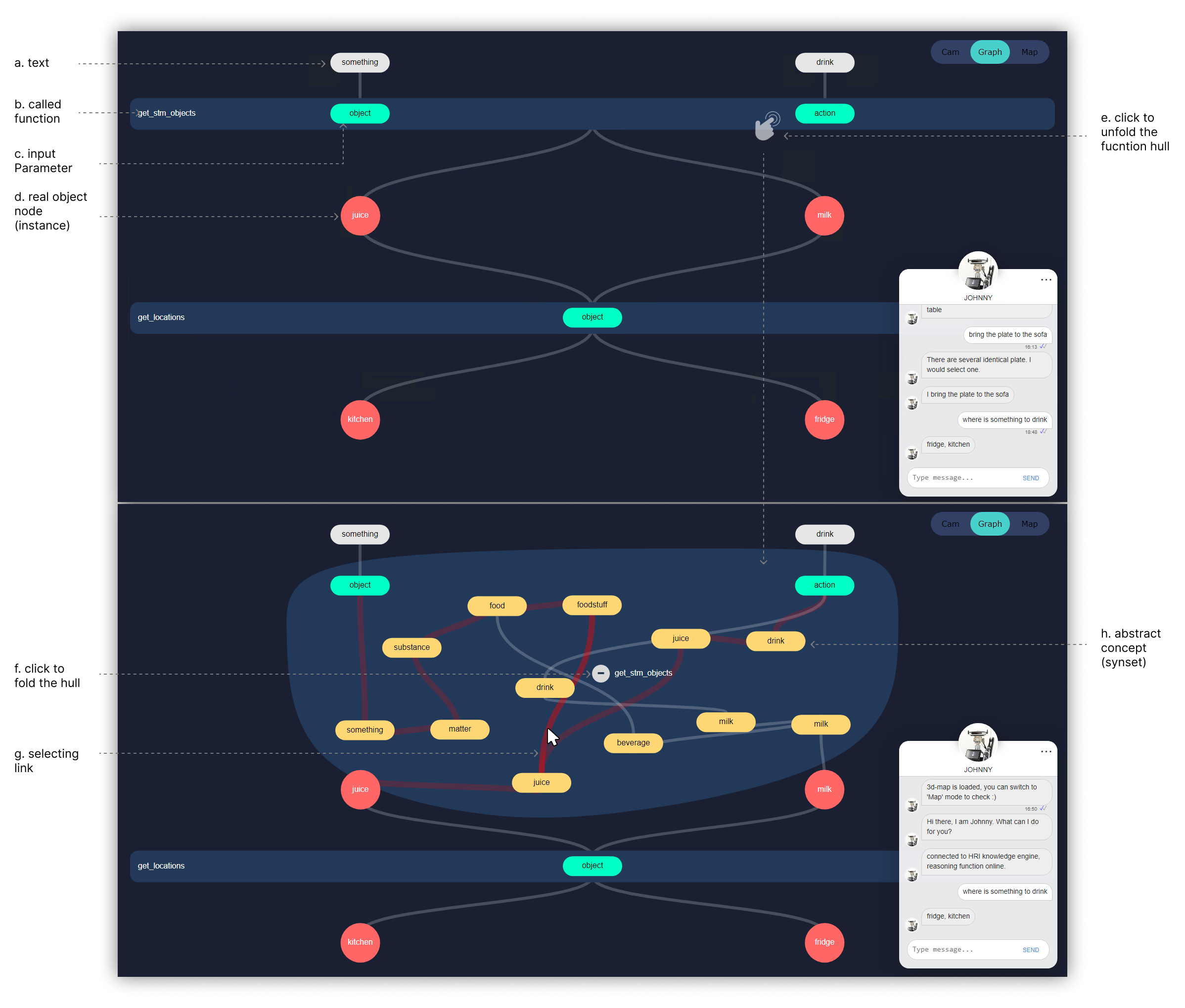}
\caption{The graph mode showing the order of executed functions (top) and, in an extended view, the connected data and paths included in the reasoning.}
\centering
\label{fig:graph-mode}
\end{figure}
The graph mode visualizes the called sequence of functions and the corresponding input/output. This is already a first step into requirement 2 and demonstrates the robot queries to the KE, based on the language input and the final output after all functions have been processed.
As shown in Figure \ref{fig:graph-mode}, after the user asks "where is something to drink?", the Semantic Parsing component converts the sentence into the "get\_stm\_objects" function (Figure \ref{fig:graph-mode}.b) in the first level. Furthermore, the module also extracts the required inputs: "something" (white node, Figure \ref{fig:graph-mode}.a) as "object" argument (green node, Figure \ref{fig:graph-mode}.c) and "drink" as "action" argument. The output of the first level function is visualized below the function bar (shown as red nodes, Figure \ref{fig:graph-mode}.d). As we have a nested function call in this example, the function results are at the same time the input of the next called function "get\_location". This function reasons about the location of the object in the environment. Finally, the final results are the "kitchen", "fridge" and "dinner table", which are the locations of objects that match with "something to drink". \par

If the user clicks the bar of the first level function, it will expand into a hull and show details about the knowledge graph structure (Figure \ref{fig:graph-mode}.e), as simplified excerpt of the actual raw data. 
The yellow nodes is abstract knowledge in the KE, which also groups synonyms that express the same concept. The lines from the utterances (white nodes) and parameters (green nodes) to the yellow node (Figure \ref{fig:graph-mode}.h) indicate that the system found the concepts which have the corresponding utterance attached. As in the user's question, the word "to drink" describes the noun "something", and the "get\_stm\_objects" function tries to find the parent concepts for "drink" and "something" and if they are related along a certain path in the knowledge graph. This mechanism of searching is visualized: The highlighted red links demonstrate that the instance object juice (red circle) is an instance of the abstract concept juice, which is the child of "foodstuff", "food", "substance", "matter" and finally "something"; and it is also connected from "juice" further to "drink". \par

As the graph data in our knowledge engine is extracted from commonsense knowledge, therefore it may contain errors, which cause problems in a robotic setting, if it cannot be corrected. According to the requirement 5 in Section \ref{sec:requirements}, in the graph mode, people can select and highlight the links for inspection. Furthermore, by pressing the "delete" button on the keyboard, the highlighted links will be deleted from the database. This changes the results of the answers by excluding the selected knowledge from reasoning.

The graph is implemented in a force simulation style of d3.js \footnote{https://github.com/d3/d3-force/tree/v3.0.0}. All the yellow nodes are dragged from a central gravity point which is the folding-button (\ref{fig:graph-mode}.f) position, which makes them clustered; At the same time, each of the nodes also has a collider to push them away from overlapping. Users can freely drag each node to put it in a preferred position. But after the user drop the node, its position will be updated by the gravity and colliders.

\subsubsection{Map mode}

\begin{figure}[h]
\includegraphics[width=0.9\textwidth]{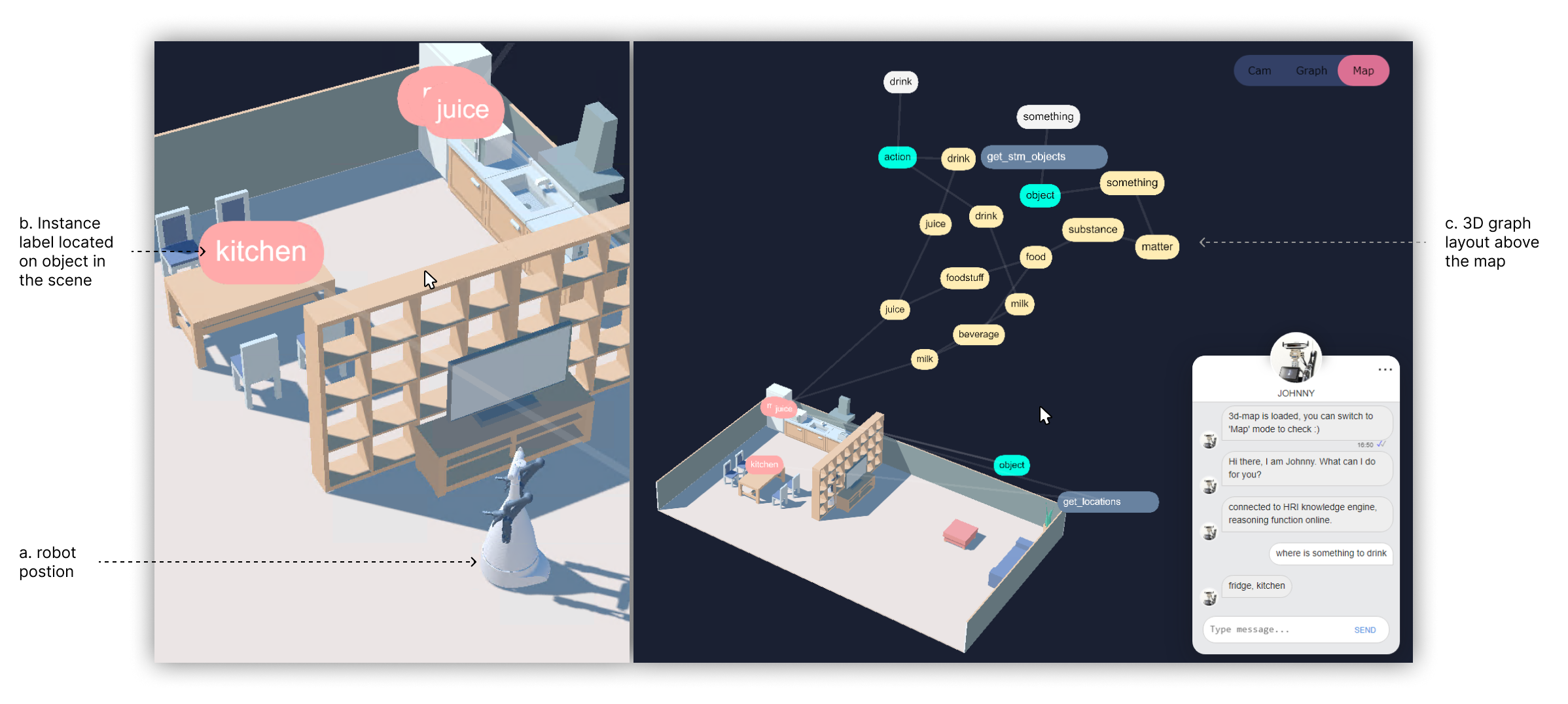}
\caption{The map mode. Right: the 3D map and 3D graph visualization. Left: user can zoom in and rotate the view to observe the map.}
\centering
\label{fig:map-mode}
\end{figure}
The map mode shows a 3D map of the simulator environment to provide a holistic view for linking the abstract knowledge graph and the scene in which the robot is operated. The 3D map contains the models of all the static objects in the scenario in a low-poly \footnote{https://en.wikipedia.org/wiki/Low\_poly} style, which ignores the details of the real objects, for providing a more concise view to the users. Above the map, a 3D graph visualization is provided. The 3D graph has exactly the same nodes and links with the 2D graph in the graph mode. But the instance node is located according to the real object's position in the 3D map (\ref{fig:map-mode}.b). In the map mode, the user can zoom in/out and rotate the view by mouse to observe the map and the 3D graph.

\section{User feedback and reflections}\label{sec:feedback}
This interface was implemented in the lab environment at our research lab and used 5 robotic experts (males, 35-45 years old) in their daily development on the system. After 3 weeks of usage, a number of insights were collected from the interview among the expert users. The developer observations and the suggestions have been collected as Gitlab issues \footnote{https://about.gitlab.com/} on a voluntary basis. In the following sections, we summarized the insight and reflections. 

\subsection{Finding and deleting the wrong links}
One of the most important features of the interface is finding the wrong links in the database. As reported by one user, after he tried to ask "where is something to eat?" the robot answered that "fork" and "salmon" can be found on the "dining room table" and in the "fridge" correspondingly. However, forks are not food. After expanding the hull in the graph mode, the user traced from the fork instance back to the eating action. Then he found there is a link between "eat" and "fork", which indicated that fork was seen as eatable in the knowledge graph. This mistake was directly corrected by deleting this link in action, and the fork was not referred to afterwards. \par
\textbf{Insights and suggestions:} The deleting feature was highly appreciated by the developers, however, they also asked to improve the graph visualization and the interactions. Firstly, there are many nodes spread randomly after the hull expanded, even though they are attracted by the centre gravity and collide with each other without overlapping, the graph still looks chaotic if too many nodes are returned. In that case, the user needs to manually arrange the position of the nodes. It is suggested to put the nodes in different layers according to the parent-child hierarchy. 
Secondly, users suggested the feature of highlighting related links, when the instance node is clicked. 
Thirdly, they also suggested to have a confirmation button or recovery button after a certain link is selected for deleting. Sometimes, either the user accidentally deletes wrong links, or they want to check how the deleted link influences the answer of the robot.
\subsection{Referred objects and attributes}
Applying a knowledge representation on a robotic system requires the abstract concept to be linked to the concrete object instance in the environment. Our interface presents this kind of grounding by highlighting a related object in the camera mode and showing the position of the object in the 3D map. A user reported these features helped them find the mismatch between the simulator environment and the knowledge graph regarding some objects' attributes. 
Another example is that one user asked for "bring the key to the kitchen". then the robot executed this command. Strangely, when the user asks "where is the key?", the robot answers 2 positions. 
After checking the map and asking for the number of the key in the kitchen and living room,
the user found the bug of the knowledge engine that it duplicated certain instance object after the state (position) changed. \par
\textbf{Insights and suggestions:} Users found that highlighting objects and positioning instance nodes in the 3D map is useful for locating referred objects. Besides this, users suggested fusing up to some extend the camera view and graph view, to reduce too many switching beteen view. Another feedback is that the chat-box which allowed users to ask questions forth and back is extremely important for identifying issues. It made the developer feel like debugging the system via natural language, which increases the efficiency and enhances the experience. At least the identification on high-level helped to find an entry point much faster, even though they might have to go back to the console or database for in-depth checking the problems sometimes.

\subsection{Visualization of called function sequence}
The feature of displaying called function sequences was welcomed by the users, as it provides a clear view of the function name, parameter and input/output, which can be easily understood by the experts. One user reported he found the bugs of the wrong-called function by reading the function name on the hull. Three users reported they found the problem that certain functions could not return correct output with the help of the visualization. \par 

\textbf{Insights and suggestions:} The visualization of the called functions explicitly demonstrates the system's decision-making process, which got high acceptance among the developers. But some improvements were suggested: sometimes the problems come from the semantic parsing component, which may fail to translate the natural languages into correct functions to call. Some users also want to see this procedure; and currently, the utterance of the output is shown, it would be better to also show concept IDs for further inspection in the knowledge graph. 

\section{Conclusions and future works}\label{sec:conlusion}
Nowadays, an increasing number of researchers utilize graph representation to enhance the capability of robots at the reasoning level. 
However, the advantage of graph representations in the perspective of transparency and explainability has not been well reflected in the current HRI interfaces and more design exploration for real-time and contextual, tailored tracing is an urgent need. 
To address this opportunity, we designed a processing-centric interface that allows the robotic expert to interact with the robot, make sense of the decision process and even modify the knowledge presented on the fly.
After being implemented in the lab and used by expert users in their daily work, they found the interface useful in supporting them to resolve problems in the system and database. 
The next step of this research is to improve the interface according to the feedback from users and to conduct a long-term research for more qualitative results. As new features, we like to tackle items 5 and 6, as listed in Section \ref{sec:requirements}, by the ability to further modify the graph by also deleting, inserting or merging action patterns, as well as inserting new action patterns or concepts via speech.

\bibliographystyle{ACM-Reference-Format}
\bibliography{CHI-2023-CaseStudy}


\begin{thebibliography}{17}


\ifx \showCODEN    \undefined \def \showCODEN     #1{\unskip}     \fi
\ifx \showDOI      \undefined \def \showDOI       #1{#1}\fi
\ifx \showISBNx    \undefined \def \showISBNx     #1{\unskip}     \fi
\ifx \showISBNxiii \undefined \def \showISBNxiii  #1{\unskip}     \fi
\ifx \showISSN     \undefined \def \showISSN      #1{\unskip}     \fi
\ifx \showLCCN     \undefined \def \showLCCN      #1{\unskip}     \fi
\ifx \shownote     \undefined \def \shownote      #1{#1}          \fi
\ifx \showarticletitle \undefined \def \showarticletitle #1{#1}   \fi
\ifx \showURL      \undefined \def \showURL       {\relax}        \fi
\providecommand\bibfield[2]{#2}
\providecommand\bibinfo[2]{#2}
\providecommand\natexlab[1]{#1}
\providecommand\showeprint[2][]{arXiv:#2}

\bibitem[F(2020)]%
        {LeuceF2020}
\bibfield{author}{\bibinfo{person}{Leuce F}.} \bibinfo{year}{2020}\natexlab{}.
\newblock \showarticletitle{On the Role of Knowledge Graphs in Explainable AI}.
\newblock \bibinfo{journal}{\emph{Semantic Web}}  \bibinfo{volume}{1}
  (\bibinfo{year}{2020}), \bibinfo{pages}{1--5}.
\newblock
Issue 0.
\urldef\tempurl%
\url{http://semantic-web-journal.org/system/files/swj2259.pdf}
\showURL{%
\tempurl}


\bibitem[Govindaraj et~al\mbox{.}(2013)]%
        {Govindaraj2013}
\bibfield{author}{\bibinfo{person}{Shashank Govindaraj},
  \bibinfo{person}{Keshav Chintamani}, \bibinfo{person}{Jeremi Gancet},
  \bibinfo{person}{Pierre Letier}, \bibinfo{person}{Boris van Lierde},
  \bibinfo{person}{Yashodhan Nevatia}, \bibinfo{person}{Geert~De Cubber},
  \bibinfo{person}{Daniel Serrano}, \bibinfo{person}{Miguel~Esbri Palomares},
  {and} \bibinfo{person}{Janusz Bedkowski}.} \bibinfo{year}{2013}\natexlab{}.
\newblock \showarticletitle{The icarus project-command, control and
  intelligence (c2i)}.
\newblock \bibinfo{journal}{\emph{2013 IEEE International Symposium on Safety,
  Security, and Rescue Robotics (SSRR)}}, \bibinfo{pages}{1--4}.
\newblock
\showISBNx{1479908800}


\bibitem[Heer et~al\mbox{.}(2005)]%
        {Heer2005}
\bibfield{author}{\bibinfo{person}{Jeffrey Heer}, \bibinfo{person}{Stuart~K
  Card}, {and} \bibinfo{person}{James~A Landay}.}
  \bibinfo{year}{2005}\natexlab{}.
\newblock \bibinfo{title}{prefuse: A Toolkit for Interactive Information
  Visualization}.
\newblock
\newblock


\bibitem[Holten and Wijk(2009)]%
        {Holten2009}
\bibfield{author}{\bibinfo{person}{Danny Holten} {and} \bibinfo{person}{Jarke
  J.~Van Wijk}.} \bibinfo{year}{2009}\natexlab{}.
\newblock \showarticletitle{Force-Directed edge bundling for graph
  visualization}.
\newblock \bibinfo{journal}{\emph{Computer Graphics Forum}}
  \bibinfo{volume}{28} (\bibinfo{year}{2009}), \bibinfo{pages}{983--990}.
\newblock
Issue 3.
\showISSN{14678659}
\urldef\tempurl%
\url{https://doi.org/10.1111/j.1467-8659.2009.01450.x}
\showDOI{\tempurl}


\bibitem[Kawamura et~al\mbox{.}(2019)]%
        {Kawamura2019}
\bibfield{author}{\bibinfo{person}{Takahiro Kawamura}, \bibinfo{person}{Shusaku
  Egami}, \bibinfo{person}{Koutarou Tamura}, \bibinfo{person}{Yasunori
  Hokazono}, \bibinfo{person}{Takanori Ugai}, \bibinfo{person}{Yusuke
  Koyanagi}, \bibinfo{person}{Fumihito Nishino}, \bibinfo{person}{Seiji
  Okajima}, \bibinfo{person}{Katsuhiko Murakami}, \bibinfo{person}{Kunihiko
  Takamatsu}, \bibinfo{person}{Aoi Sugiura}, \bibinfo{person}{Shun Shiramatsu},
  \bibinfo{person}{Shawn Zhang}, {and} \bibinfo{person}{Kouji Kozaki}.}
  \bibinfo{year}{2019}\natexlab{}.
\newblock \showarticletitle{Report on the first knowledge graph reasoning
  challenge 2018 - toward the eXplainable AI system -}.
\newblock \bibinfo{journal}{\emph{arXiv}} (\bibinfo{year}{2019}),
  \bibinfo{pages}{1--18}.
\newblock
\showISSN{23318422}


\bibitem[Ma et~al\mbox{.}(2019)]%
        {Ma2019}
\bibfield{author}{\bibinfo{person}{Weizhi Ma}, \bibinfo{person}{Woojeong Jin},
  \bibinfo{person}{Min Zhang}, \bibinfo{person}{Chenyang Wang},
  \bibinfo{person}{Yue Cao}, \bibinfo{person}{Yiqun Liu},
  \bibinfo{person}{Shaoping Ma}, {and} \bibinfo{person}{Xiang Ren}.}
  \bibinfo{year}{2019}\natexlab{}.
\newblock \showarticletitle{Jointly learning explainable rules for
  recommendation with knowledge graph}.
\newblock \bibinfo{journal}{\emph{The Web Conference 2019 - Proceedings of the
  World Wide Web Conference, WWW 2019}} (\bibinfo{year}{2019}),
  \bibinfo{pages}{1210--1221}.
\newblock
\showISBNx{9781450366748}
\urldef\tempurl%
\url{https://doi.org/10.1145/3308558.3313607}
\showDOI{\tempurl}


\bibitem[Nielsen and Goodrich(2006)]%
        {Nielsen2006}
\bibfield{author}{\bibinfo{person}{Curtis~W. Nielsen} {and}
  \bibinfo{person}{Michael~A. Goodrich}.} \bibinfo{year}{2006}\natexlab{}.
\newblock \showarticletitle{Comparing the usefulness of video and map
  information in navigation task}.
\newblock \bibinfo{journal}{\emph{HRI 2006: Proceedings of the 2006 ACM
  Conference on Human-Robot Interaction}}  \bibinfo{volume}{2006}
  (\bibinfo{year}{2006}), \bibinfo{pages}{95--101}.
\newblock
\showISBNx{1595932941}
\urldef\tempurl%
\url{https://doi.org/10.1145/1121241.1121259}
\showDOI{\tempurl}


\bibitem[Puig et~al\mbox{.}(2018)]%
        {Puig2018}
\bibfield{author}{\bibinfo{person}{Xavier Puig}, \bibinfo{person}{Kevin Ra},
  \bibinfo{person}{Marko Boben}, \bibinfo{person}{Jiaman Li},
  \bibinfo{person}{Tingwu Wang}, \bibinfo{person}{Sanja Fidler}, {and}
  \bibinfo{person}{Antonio Torralba}.} \bibinfo{year}{2018}\natexlab{}.
\newblock \showarticletitle{Virtualhome: Simulating household activities via
  programs}.
\newblock \bibinfo{journal}{\emph{Proceedings of the IEEE Conference on
  Computer Vision and Pattern Recognition}}, \bibinfo{pages}{8494--8502}.
\newblock


\bibitem[Strobelt et~al\mbox{.}(2019)]%
        {Strobelt2019}
\bibfield{author}{\bibinfo{person}{Hendrik Strobelt},
  \bibinfo{person}{Sebastian Gehrmann}, \bibinfo{person}{Michael Behrisch},
  \bibinfo{person}{Adam Perer}, \bibinfo{person}{Hanspeter Pfister}, {and}
  \bibinfo{person}{Alexander~M. Rush}.} \bibinfo{year}{2019}\natexlab{}.
\newblock \showarticletitle{Seq2seq-Vis: A Visual Debugging Tool for
  Sequence-to-Sequence Models}.
\newblock \bibinfo{journal}{\emph{IEEE Transactions on Visualization and
  Computer Graphics}}  \bibinfo{volume}{25} (\bibinfo{year}{2019}),
  \bibinfo{pages}{353--363}.
\newblock
Issue 1.
\showISSN{19410506}
\urldef\tempurl%
\url{https://doi.org/10.1109/TVCG.2018.2865044}
\showDOI{\tempurl}


\bibitem[Szafir et~al\mbox{.}(2017)]%
        {Szafir2017}
\bibfield{author}{\bibinfo{person}{Daniel Szafir}, \bibinfo{person}{Bilge
  Mutlu}, {and} \bibinfo{person}{Terrence Fong}.}
  \bibinfo{year}{2017}\natexlab{}.
\newblock \showarticletitle{Designing planning and control interfaces to
  support user collaboration with flying robots}.
\newblock \bibinfo{journal}{\emph{International Journal of Robotics Research}}
  \bibinfo{volume}{36} (\bibinfo{date}{6} \bibinfo{year}{2017}),
  \bibinfo{pages}{514--542}.
\newblock
Issue 5-7.
\showISSN{17413176}
\urldef\tempurl%
\url{https://doi.org/10.1177/0278364916688256}
\showDOI{\tempurl}


\bibitem[Szafr and Szafr(2021)]%
        {Szafr2021}
\bibfield{author}{\bibinfo{person}{Daniel Szafr} {and}
  \bibinfo{person}{Danielle~Albers Szafr}.} \bibinfo{year}{2021}\natexlab{}.
\newblock \showarticletitle{Connecting human-robot interaction and data
  visualization}.
\newblock \bibinfo{journal}{\emph{ACM/IEEE International Conference on
  Human-Robot Interaction}}, \bibinfo{pages}{281--292}.
\newblock
\showISBNx{9781450382892}
\showISSN{21672148}
\urldef\tempurl%
\url{https://doi.org/10.1145/3434073.3444683}
\showDOI{\tempurl}


\bibitem[Tiddi and Schlobach(2022)]%
        {Tiddi2022}
\bibfield{author}{\bibinfo{person}{Ilaria Tiddi} {and} \bibinfo{person}{Stefan
  Schlobach}.} \bibinfo{year}{2022}\natexlab{}.
\newblock \showarticletitle{Knowledge graphs as tools for explainable machine
  learning: A survey}.
\newblock \bibinfo{journal}{\emph{Artificial Intelligence}}
  \bibinfo{volume}{302} (\bibinfo{date}{1} \bibinfo{year}{2022}),
  \bibinfo{pages}{103627}.
\newblock
\showISSN{00043702}
\urldef\tempurl%
\url{https://doi.org/10.1016/j.artint.2021.103627}
\showDOI{\tempurl}


\bibitem[Wang et~al\mbox{.}(2018)]%
        {Wang2018}
\bibfield{author}{\bibinfo{person}{Xiang Wang}, \bibinfo{person}{Dingxian
  Wang}, \bibinfo{person}{Canran Xu}, \bibinfo{person}{Xiangnan He},
  \bibinfo{person}{Yixin Cao}, {and} \bibinfo{person}{Tat~Seng Chua}.}
  \bibinfo{year}{2018}\natexlab{}.
\newblock \showarticletitle{Explainable reasoning over knowledge graphs for
  recommendation}.
\newblock \bibinfo{journal}{\emph{arXiv}}  \bibinfo{volume}{33},
  \bibinfo{pages}{5329--5336}.
\newblock
Issue 01.
\showISBNx{2374-3468}
\showISSN{23318422}


\bibitem[Wattenberg(2002)]%
        {Wattenberg2002}
\bibfield{author}{\bibinfo{person}{M. Wattenberg}.}
  \bibinfo{year}{2002}\natexlab{}.
\newblock \showarticletitle{Arc diagrams: Visualizing structure in strings}.
\newblock \bibinfo{journal}{\emph{Proceedings - IEEE Symposium on Information
  Visualization, INFO VIS}}  \bibinfo{volume}{2002-January},
  \bibinfo{pages}{110--116}.
\newblock
\showISBNx{076951751X}
\showISSN{1522404X}
\urldef\tempurl%
\url{https://doi.org/10.1109/INFVIS.2002.1173155}
\showDOI{\tempurl}


\bibitem[Whitlock et~al\mbox{.}(2020)]%
        {Whitlock2020}
\bibfield{author}{\bibinfo{person}{Matt Whitlock}, \bibinfo{person}{Keke Wu},
  {and} \bibinfo{person}{Danielle~Albers Szafir}.}
  \bibinfo{year}{2020}\natexlab{}.
\newblock \showarticletitle{Designing for Mobile and Immersive Visual Analytics
  in the Field}.
\newblock \bibinfo{journal}{\emph{IEEE Transactions on Visualization and
  Computer Graphics}}  \bibinfo{volume}{26} (\bibinfo{year}{2020}),
  \bibinfo{pages}{503--513}.
\newblock
Issue 1.
\showISSN{19410506}
\urldef\tempurl%
\url{https://doi.org/10.1109/TVCG.2019.2934282}
\showDOI{\tempurl}


\bibitem[Xian et~al\mbox{.}(2019)]%
        {Xian2019}
\bibfield{author}{\bibinfo{person}{Yikun Xian}, \bibinfo{person}{Zuohui Fu},
  \bibinfo{person}{S. Muthukrishnan}, \bibinfo{person}{Gerard~De Melo}, {and}
  \bibinfo{person}{Yongfeng Zhang}.} \bibinfo{year}{2019}\natexlab{}.
\newblock \showarticletitle{Reinforcement knowledge graph reasoning for
  explainable recommendation}.
\newblock \bibinfo{journal}{\emph{SIGIR 2019 - Proceedings of the 42nd
  International ACM SIGIR Conference on Research and Development in Information
  Retrieval}} (\bibinfo{year}{2019}), \bibinfo{pages}{285--294}.
\newblock
\showISBNx{9781450361729}
\urldef\tempurl%
\url{https://doi.org/10.1145/3331184.3331203}
\showDOI{\tempurl}


\bibitem[Yanco et~al\mbox{.}(2007)]%
        {Yanco2007}
\bibfield{author}{\bibinfo{person}{Holly~A. Yanco}, \bibinfo{person}{Brenden
  Keyes}, \bibinfo{person}{Jill~L. Drury}, \bibinfo{person}{Curtis~W. Nielsen},
  \bibinfo{person}{Douglas~A. Few}, {and} \bibinfo{person}{David~J. Bruemmer}.}
  \bibinfo{year}{2007}\natexlab{}.
\newblock \showarticletitle{Evolving interface design for robot search tasks}.
\newblock \bibinfo{journal}{\emph{Journal of Field Robotics}}
  \bibinfo{volume}{24}, \bibinfo{pages}{779--799}.
\newblock
Issue 8-9.
\showISSN{15564959}
\urldef\tempurl%
\url{https://doi.org/10.1002/rob.20215}
\showDOI{\tempurl}


\end{thebibliography}

\end{document}